          [1999/12/02 1.01 (PWD)]
\documentclass[a4paper,twocolumn]{esapub}
\usepackage{graphics}

\usepackage{natbib}
\newcommand{\be}{\begin{equation}}
\newcommand{\ee}{\end{equation}}
\title{On the nature of the unidentified MeV gamma-ray source GRO J1411-64}

\author{Gustavo E. Romero}
\author{Mariana Orellana}
\affil{\it IAR, C.C. 5, 1894 Villa Elisa, Argentina. E-mail: romero@venus.fisica.unlp.edu.ar}
\affil{\it Facultad de Ciencias Astron\'omicas y Geof\'{\i}sicas, UNLP, Paseo del Bosque, 1900 La Plata, Argentina.}
\author{Diego F. Torres}
\affil{\it Lawrence
Livermore National Laboratory, 7000 East Ave., L-413, Livermore,
CA 94550, USA.}

\begin{document}


\maketitle

\begin{abstract}
A recent analysis of COMPTEL data has revealed the existence of an unidentified and variable gamma-ray source, designated GRO J1411-64, at MeV energies in the galactic plane (Zhang et al. 2002). We discuss the possibility that such a source can be produced in the accreting neutron star system 2S 1417-624 through the bombardment of the accretion disk by relativistic protons accelerated in the magnetosphere by the Cheng-Ruderman mechanism. 
\end{abstract}

\section{Introduction}
 
 In a recent communication, Zhang et al. (2002) have reported the discovery of a new source near the galactic plane in the COMPTEL data of the Compton Gamma-Ray Observatory (CGRO). The source is detected at 7.2$\sigma$ level in a combination of 7 viewing periods (VP) of the satellite, during 1995. The total effective exposure during Phases 1 to 5 of CGRO was of more than 35 days. The new source is clearly variable in the 1-3 MeV band on timescales of months. The maximum flux in this energy range is $\sim 20\;10^{-5}$ ph cm$^{-2}$ s$^{-1}$($\sim 0.3$ Crab). The spectrum can be represented by a simple power law with a photon index $\Gamma=2.61^{+0.33}_{-0.30}$. 
 
 The clear variability of the source indicates that it should be a compact object. Zhang et al. (2002) have searched for possible counterparts within the error location contours of the COMPTEL detection. There are several interesting objects in this region of the sky, including two high-velocity clouds (HVC), two radio pulsars, two Wolf-Rayet (WR) stars, one transient black hole binary, and a Be/X-ray transient source. 
The variability and the soft MeV spectrum of the source rules out the pulsars, which are typically non-variable and hard sources, and strongly constrain the presumed emission region in the HVC. The WR stars are potential MeV emitters if they are colliding wind binaries. Benaglia \& Romero (2003) discuss in detail how these kind of stars can generate MeV sources when they are in binary systems. In the present case, however, there is no evidence for binarity. Variable MeV emission from  microquasars has also been recently discussed by Romero et al. (2002) in connection to Cygnus X-1, and remains as an open possibility for GRO J1411-64 (see also the microquasar models discussed by Romero et al. in these proceedings). Here, we will focus on the possible gamma-ray emission from accreting neutron star systems like 2S 1417-624, which is the Be/X-ray transient source within the COMPTEL error location contours, in an attempt to establish whether this is a likely counterpart.        


\section{Gamma-ray production in accreting neutron stars}

Be/X-ray transients are binary systems formed by a massive Be star and a magnetized neutron star in an eccentric orbit. Be stars are rapidly rotating objects with strong equatorial winds. The neutron star can accrete directly from the circumstellar Be wind to form a transient accretion disk during a periastron passage. Such disks have been found in systems like A0535+26 through the detection of quasi-periodic oscillations in the power spectrum of the X-ray flux (e.g. Finger et al. 1996a). The parameters that characterize the disk will change as the neutron star moves along the orbit. Eventually, at some distance that will depend on the specific system considered, there might be a change in the accretion regime from disk accretion to Bondi-Hoyle accretion.

Once formed, the accretion disk can penetrate the magnetosphere of the rotating neutron star (Gosh \& Lamb 1979). This penetration creates a broad transition region between the unperturbed disc flow and the co-rotating magnetospheric flow. In this transition zone, at the inner radius $R_0$ the angular velocity is still Keplerian. Between $R_0$ and the co-rotation radius $R_{\rm co}$ there is a thin boundary layer where the angular velocity departs from the Keplerian value. At $R_{\rm co}$ the disc is disrupted by the magnetic pressure and the matter is channeled by the magnetic field lines to the neutron star polar cap, where it impacts producing hard X-ray emission.

\begin{figure}[t]
\centering
\begin{flushleft}
\resizebox{8cm}{!} {\includegraphics{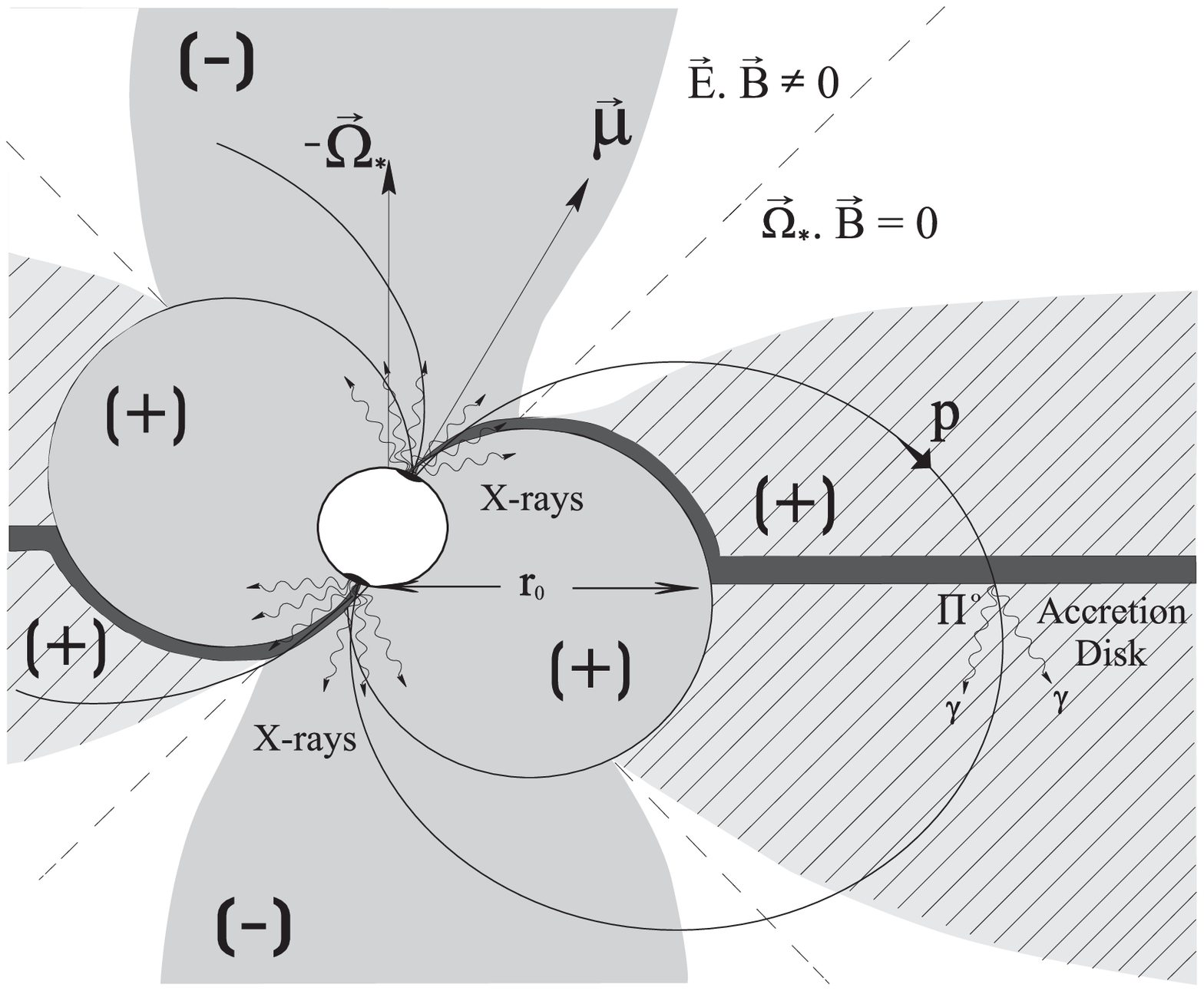}}
\caption{Sketch of the magnetosphere of an accreting magnetized neutron star. An electrostatic gap is produced when $\Omega_*<\Omega_{\rm disk}$. Protons can be accelerated in this gap along the magnetic field lines up to multi-TeV energies before impacting onto the disk, where they produce high-energy gamma-rays through $\pi^0$-decays. These gamma-rays, in turn, can initiate electromagnetic cascades in the disk. The emerging gamma-rays, on the other side of the disk, can be also degraded in the X-ray photosphere to MeV energies before reaching the observer.\label{fig2}}
\end{flushleft}
\end{figure}

Cheng \& Ruderman (1989, 1991) have studied the magnetosphere of accreting neutron stars when the disk rotates faster than the star, showing that three definite regions are formed: 1. A region coupled to the star by the magnetic field lines that do not penetrate the disk, 2. An equatorial region linked to the disk by the field attached to it, and 3. A gap entirely empty of plasma separating both regions. In this gap ${\bf E\cdot B}\neq 0$ and a strong potential drop $\Delta V$ is established (see Fig. \ref{fig2} for a sketch of the situation). For typical parameters in Be/X-ray binaries, $\Delta V\sim 10^{13-14}$ V (Romero et al. 2001). Protons entering into the gap from the stellar co-rotating region will be accelerated to multi-TeV energies and directed to the accretion disk by the field lines, where they will impact producing $\pi^{\pm}$ and $\pi^0$ in $pp$ interactions with the disk material.  These pions will decay inside the disk producing electron-positron pairs, high-energy (TeV) photons, and neutrinos. The neutrinos will escape from the system (see Anchordoqui et al. 2003 for estimates of the neutrino flux), but the pairs and the gamma-rays initiate electromagnetic cascades that will degrade the energy of the gamma-ray photons to form a standard cascade spectrum with a cut-off at MeV energies. The leptons in the disk will lose energy by inverse Compton (IC) interactions with the thermal X-ray photons, by synchrotron radiation in the field that penetrates the disk, and by relativistic Bremsstrahlung in the ions of the accreting gas. The details of the cascade development will be determined by the relative ratios of the cooling times among all these processes and the opacities for the produced photons in the specific cases.    

\section{The system 2S 1417-64}

The X-ray source 2S 1417-64 was discovered by Apparao et al. (1980) using the SAS-3 satellite. Fourier analysis of these data revealed the presence of pulsations with a period of $\sim17.63$ sec. Observations with the Einstein satellite and optical telescopes led to the detection of a Be star companion (Grindlay et al. 1984). The distance to the system is not well determined due to the uncertainties in the spectral type; it might be in the range 1.4--11.1 kpc. The source was detected by BATSE (Finger et al. 1996b) and by RXTE ($\rm \dot{I}$nam et al. 2004).  Large outbursts followed by a series of mini-bursts were reported, a behavior similar to what is observed in A0535+26. The orbital period and the eccentricity of the system are $P=42.19$ days and $e=0.446$, respectively ($\rm \dot{I}$nam et al. 2004). The X-ray data are consistent with the formation of a transient accretion disk during the outbursts. The larger outbursts last about two orbital periods, starting immediately after the periastron passage. During these outbursts the circumstellar disk of the Be star should expand in such a way that it can provide matter to the neutron star to form a transient accretion disk when the periastron interaction occurs. The density of the circumstellar disk diminishes during the mini-outbursts, in such a way that the accretion rate is smaller and the accreting material is replenished with each passage until the disk retracts completely. The soft, unpulsed X-ray emission is originated in the thermal disk, whereas the pulsed hard X-rays are from matter impacting onto the neutron star polar cap.         

\section{Specific models}

In order to study the gamma-ray production in 2S 1417-62 during the accretion disk phase, we adopt the following values for the basic parameters of the system: $M_{\rm NS}=1.4$ $M_{\odot}$, $R_{\rm NS}=10^6$ cm, $B=10^{12}$ G, and $\eta=0.2$, which correspond to the neutron star mass, radius, surface dipole magnetic field, and screening factor of the field in the disk ($\eta=0$ corresponds to a perfectly diamagnetic disk), respectively. With these parameters and the X-ray history of the major outburst reported by Finger et al. (1996b) we calculate the structure of the accretion disk (Shakura \& Sunyaev 1973, Gosh \& Lamb 1979) and the magnetosphere (Cheng \& Ruderman 1991) as a function of time. Since the distance to the source is unknown, we calculated three different models corresponding to peak luminosities of the major outbursts of $10^{37}$, $5\;10^{36}$ and $10^{36}$ erg s$^{-1}$. For each model we then estimated the proton current impacting onto the accretion disk surface, taking into account photo-pion and curvature radiation losses during the acceleration process (the proton losses resulted negligible in all models). Then we calculated, for different times of the outburst, the averaged pair cascades initiated in the disk and the opacities for the propagation of the emerging gamma-ray photons in the ambient X-ray fields once they escape from the disk. These results are briefly discussed in the next section.        

\section{Results}

\begin{figure}[t]
\centering
\resizebox{8cm}{!} {\includegraphics{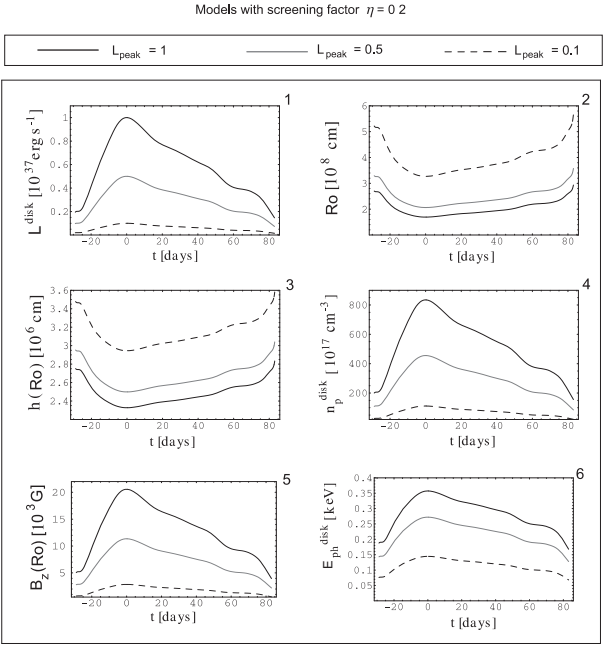}}
\caption{Evolution of disk parameters for three different models characterized by their peak luminosities. From left to right and from top to bottom: 1. Disk X-ray luminosity as a function of time ($t=0$ corresponds to the periastron passage), 2. Inner radius ($R_0$) of the accretion disk, 3. Half-height of the accretion disk at $R_0$, 4. Particle density of the disc at $R_0$, 5. Magnetic field inside the disk at $R_0$, and 6. Average energy of the disk photons at $R_0$. \label{fig3}}
\end{figure}

\begin{figure}[t]
\centering
\resizebox{8cm}{!} {\includegraphics{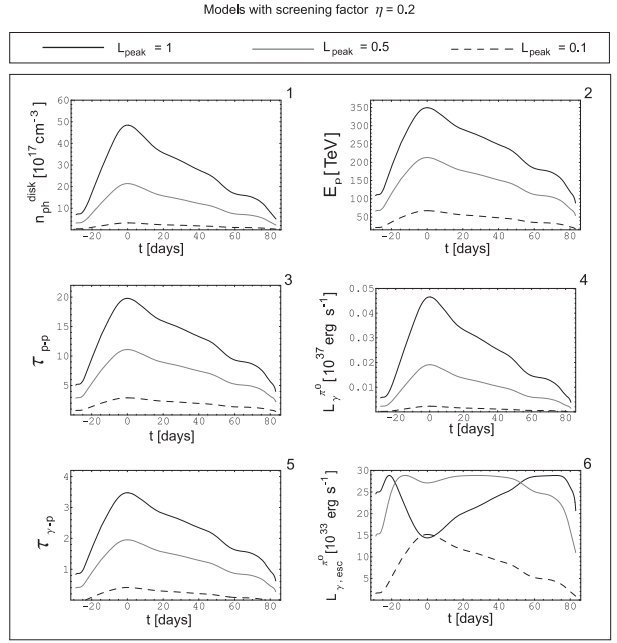}}
\caption{From left to right and from top to bottom: 1. Photon density in the disk at $R_0$ as a function of time, 2. Energy of the protons impacting onto the disk, 3. Opacity of the disk to the protons, 4. Gamma-ray luminosity produced by $\pi^0$ decays inside the disk, 5. Opacity of the disk to the $\pi^0$ gamma-rays, and 6. The emerging $\pi^0$ gamma-ray luminosity from the opposite side of the disk. \label{fig4}}
\end{figure}

In Fig. \ref{fig3} we show the main parameters that characterize the accretion disk along the evolution of a major outburst. The different curves in each plot correspond to models with different peak luminosities, namely $L_{\rm peak}=$ 1, 0.5, and 0.1, in units of $10^{37}$ erg s$^{-1}$. In Fig. \ref{fig4} we show the evolution of the photon density inside the disk and the energy of the gap-accelerated protons that impact onto the disk surface. These protons have energies in the ranges 110--350 TeV and 10--70 TeV for the most and least luminous models, respectively. Another panel of the figure shows the optical depth of the disk to proton propagation. It is clear that the disk is always opaque to the relativistic protons. The gamma-ray luminosity produced by $\pi^0$-decays can be as high as $\sim 5\;10^{35}$ erg s$^{-1}$ for the model with $L_{\rm peak}=1$. The energy of these gamma-rays will be in the range 20--60 TeV. The disk is mostly opaque to the propagation of these photons. The models  with disks of smaller column densities are partially transparent. The emerging very high energy gamma-ray flux (see last panel of the figure) can be detected in all cases by the new generation of imaging Cherenkov telescopes, for reasonable distances. The source, of course, should be observed during a major outburst.

\begin{figure}[t]
\centering
\resizebox{8cm}{!} {\includegraphics{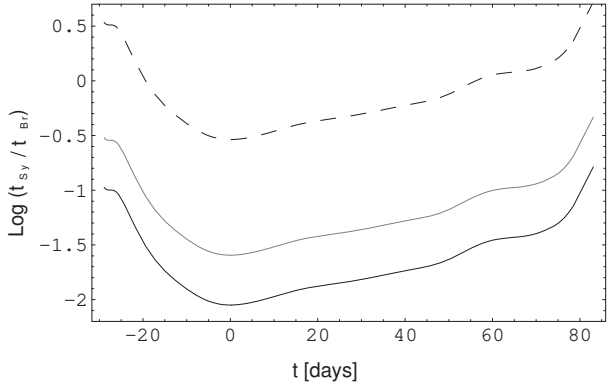}}
\caption{Ratio of synchrotron to Bremsstrahlung cooling times for the first generation of $e^{\pm}$-pairs injected into the accretion disk.\label{fig5}}
\end{figure}

\begin{figure}[t]
\centering
\resizebox{8cm}{!}{\includegraphics{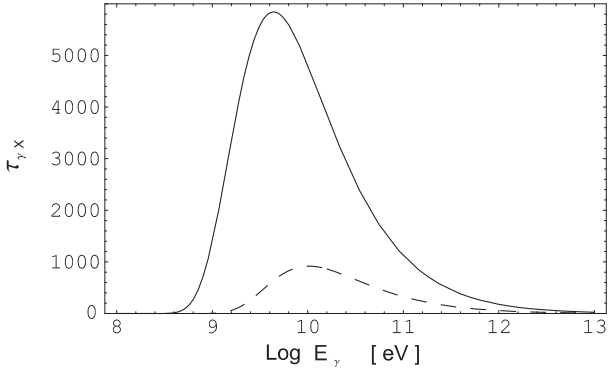}}
\caption{Opacity of the X-ray photosphere to gamma-ray propagation at $t=0$ for $L_{\rm peak}=1$ and $0.1$, as a function of the gamma-ray energy. \label{fig6}}
\end{figure}

Those gamma-rays that are absorbed in the disk by photon-photon pair creation initiate electromagnetic cascades. Cascades are also initiated by the $e^{\pm}$-pairs injected through the decay of $\pi^{\pm}$. The first generation of pairs mainly cools by synchrotron and Bremsstrahlung radiation. In Fig. \ref{fig5} we show the ratio of synchrotron to Bremsstrahlung cooling times for these particles. Since the medium is still opaque to the second generation gamma-rays, there will be a second, and then a third generation of $e^{\pm}$-pairs, that will cool mainly by relativistic Bremsstrahlung. The gamma-rays that escape from the opposite side of the accretion disk will be injected into the X-ray photon field of the disk where they will be absorbed initiating an IC cascade that will degrade their energy to the soft MeV band. In Fig. \ref{fig6} we show the opacity of the X-ray field  as a function of energy for $t=0$, in the case of $L_{\rm peak}=1$. This field is the addition of the soft blackbody disk component plus a hard X-ray component from the pulsar's polar cap that we have modeled as a power-law with index $\Gamma=1$ and an exponential cut-off at $kT=200$ keV. It can be seen that the GeV emission is completely suppressed. This leads to the absence of an EGRET source in this case. Only in the model with $L_{\rm peak}=0.1$ the GeV gamma-rays can escape, and this only when the disk is close to its minimum luminosity.

The MeV photons can always escape from the system. This emission will be variable because of the evolution of all relevant parameters along the orbit. The final cascade spectrum will be soft, with an index $\Gamma\sim2.5$ and a high-energy cut-off around $\sim 2m_{\rm e}^2 c^4/ \left\langle E_{\rm ph}\right\rangle$, where $\left\langle E_{\rm ph}\right\rangle$ is the mean photon energy of the ambient fields (see Aharonian et al. 1985 for further details on cascades in accretion disks). 

We see, then, that several characteristics of the MeV gamma-ray source can be explained under the assumption that 2S 1417-64 is the counterpart: soft MeV spectrum, variability, and absence of EGRET detection. However, there are other features of the model that are difficult to reconcile with the observational data: correlation with the X-ray variability (which is expected at MeV energies, contrary to what happens at GeV energies) and a ratio $f=L_X/L_{\gamma}>1$. In fact, for all models we expect $f\sim10$, instead of the observed $f\sim1$. The value of $f$ might be modified by a different assumption of the magnetic field in the neutron star, but the dependence is not strong.  

\section{Conclusions}

Integral observations are crucial to establish the nature of the enigmatic source GRO J1411-64. Our models for the accreting neutron star 2S 1417-62 indicate that it might be a variable MeV source with a total gamma-ray luminosity of $\sim 10^{36}$ erg s$^{-1}$ when the accretion disk is fully formed in models with a disk luminosity of $10^{37}$ erg s$^{-1}$ (a weaker source is predicted by the other cases studied here). Whether the MeV flux can be detected depends on the distance to the source, which is unknown. If the suggested existence of anti-correlation between the MeV and the X-ray radiation is confirmed (Zhang et al. 2002), 2S 1417-62 might be ruled out as counterpart since this anti-correlation should exist only with the emission above 100 MeV, due to opacity effects in the photosphere. In all cases a significant TeV source is expected. Depending on the disk density this source can be highly variable or roughly steady during a major outbursts. In some models the TeV emission is anti-correlated with the X-ray luminosity, as it is shown in the last panel of Fig. \ref{fig4}. Sources of this kind might be detectable with modern Cherenkov telescopes (Cheng et al. 1991).

\subsection*{Acknowledgments}

This work has been supported by Fundaci\'on Antorchas and CONICET (PICT 0438/98). GER thanks 
the SECyT (Argentina) for a Houssay Prize.



\begin{thebibliography}{}

\bibitem{}Aharonian F.A., et al., 1985, {\it Ap\&SS} 115, 201 
\bibitem{}Anchordoqui L.A., et al., 2003, {\it ApJ} 589, 481
\bibitem{}Apparao K.M.V., et al., 1980, {\it A\&A} 89, 249
\bibitem{}Benaglia P., Romero G.E., 2003, {\it A\&A} 399, 1121 
\bibitem{}Cheng K.S., Ruderman M., 1989, {\it ApJ} 337, L77
\bibitem{}Cheng K.S., Ruderman M., 1991, {\it ApJ} 373, 187
\bibitem{}Cheng K.S., et al., 1991, {\it ApJ} 379, 290
\bibitem{}Finger M.H., et al., 1996a, {\it ApJ} 459, 288
\bibitem{}Finger M.H., et al., 1996b, {\it A\&AS} 120, 209
\bibitem{}Gosh P., Lamb F.K., 1979, {\it ApJ} 234, 296
\bibitem{}Grindlay J.E., et al., 1984, {\it ApJ} 276, 621
\bibitem{}$\rm \dot{I}$nam S.C., et al., 2004, {\it MNRAS}, in press
\bibitem{}Romero G.E., et al., 2001, {\it A\&A} 376, 599
\bibitem{}Romero G.E., et al., 2002, {\it A\&A} 393, L61 
\bibitem{}Shakura N.I., Sunyaev R.A. 1973, {\it A\&A} 24, 337
\bibitem{}Zhang S., et al., 2002, {\it A\&A} 396, 923
\end{thebibliography}
\end{document}